\documentclass[%
 reprint,
 amsmath,amssymb,
 aps,
 doi,
prd,
floatfix,
]{revtex4-2}

\usepackage{graphicx}
\usepackage{dcolumn}
\usepackage{bm}
\usepackage{multirow}
\usepackage{url}
\usepackage{orcidlink}
\usepackage{cleveref}

\begin{document}

\preprint{APS/123-QED}

\title{Atmospheric Neutrino Oscillations:\\the Full Picture}

\author{Philipp Eller\orcidlink{0000-0001-6354-5209}}
 \altaffiliation{Physics Department, Technical University of Munich}
 \email{philipp.eller@tum.de}

\date{\today}

\begin{abstract}
We present the first combined oscillation analysis of recent atmospheric neutrino 
datasets, featuring data from Super-Kamiokande, IceCube-DeepCore, and 
KM3NeT/ORCA together with reactor data from Daya Bay. Such combinations have 
long been considered infeasible outside experimental collaborations; we demonstrate 
that a unified physics model can simultaneously describe all datasets with no 
significant parameter tensions. Fitting 839\,048 events across 1536 bins with 91 
parameters, our combined analysis yields competitive measurements 
of the neutrino mixing parameters, and prefers the Normal 
over the Inverted Mass Ordering at $3\sigma$ significance.
\end{abstract}

\maketitle

\section{Introduction}

Atmospheric neutrinos have played a pivotal role in the study of neutrino oscillations from early on. The Super-K collaboration famously reported 5$\sigma$ evidence against the no-oscillation hypothesis in 1998 \cite{Super-Kamiokande:1998kpq}. The oscillation interpretation was then confirmed in a complementary channel by the SNO collaboration using solar neutrinos in 2002 \cite{SNO:2002tuh}.

Since then, a multitude of experiments have measured the three oscillation angles and the (absolute) values of the mass splitting to high precision. All in all, good agreement with the 3-flavor oscillation picture is observed, with a few notable exceptions known as "anomalies".

What remains still unknown, however, are the presence of CP-violation and the ordering of the neutrino masses (NMO). From solar oscillations, it was established that the second mass eigenstate is heavier than the first ($m_1 < m_2$) \cite{SNO:2002tuh, KamLAND:2002uet}, but it is not yet clear whether the third mass eigenstate is the heaviest ($m_1 < m_2 < m_3$; Normal Ordering (NO)), or the lightest ($m_3 < m_1 < m_2$; Inverted Ordering (IO)). 

These two questions have been studied using accelerator neutrinos, most recently by analyses from the T2K collaboration \cite{T2K:2025yoy}, favoring a value of $\delta_{CP}$ of around $3/2\pi$ and a Bayes factor of 1.7 preferring the NO over the IO (or a $\Delta\chi^2_{IO-NO}\approx 2.1$), and the NO$\nu$A collaboration \cite{NOvA:2025tmb} also reported a similar $1.4\sigma$ preference of the NO over the IO, however, their data prefers a $\delta_{CP}$ value of $0.87\pi$ assuming NO. 

It was realized, as for example discussed in \cite{Esteban:2020cvm, deSalas:2020pgw, Capozzi:2025wyn}, that this tension in the $\delta_{CP}$ phase is problematic when combining the two datasets. In fact, since the tension in $\delta_{CP}$ is not present under IO, a combined fit does prefer IO over NO (Bayes factor 1.3). Even though both individual fits prefer NO over IO. This tension between the two datasets was also confirmed by an official result in a joint analysis of both collaborations \cite{T2K:2025wet}.

Another experiment with sensitivity to the NMO that has already started taking data is JUNO, which recently reported its first oscillation results \cite{JUNO:2025gmd}. However, only solar parameters were constrained, and more data is needed to make statements on the NMO.

Independent sensitivity to the NMO comes from matter effects on Earth-crossing atmospheric neutrinos. Super-K reports a preference of NO vs. IO with a $\Delta\chi^2_{IO-NO}=5.69$ \cite{Super-Kamiokande:2023ahc}. An effort combining Super-K with T2K reports similar findings \cite{T2K:2024wfn}.
IceCube has not made any official statement about the NMO since 2019, when they reported a mild preference of NO over IO at $\Delta\chi^2_{IO-NO}=0.738$ using 3 years of data \cite{IceCube:2019dyb}. A recent PhD thesis, however, reports $\Delta\chi^2_{IO-NO}=4.398$ using 9.3 years of IceCube data \cite{maria_phd}.
KM3NeT observes a similarly mild preference of NO over IO in their first oscillation analysis of $\Delta\chi^2_{IO-NO}=0.62$ \cite{KM3NeT:2024ecf}.

The question is: can these neutrino experiments be combined into a consistent picture to measure oscillation parameters and make a combined statement on the NMO?

Such combinations of public datasets have been the tale of multiple "global fitter" groups for many years, producing very successful global results on oscillation parameters and beyond the Standard Model (BSM) physics.
The sore spot in these combinations is the lack of atmospheric data included. Some groups had expressed inability to reproduce the Super-K results, as discussed in \cite{Esteban:2016qun}, with the last inclusion of Super-K in a combined fit in 2014 \cite{Gonzalez-Garcia:2014bfa}. And only after that point in time have IceCube data been introduced into the analysis in 2018 \cite{Esteban:2018azc}, by which time Super-K had already been removed. In Ref.~\cite{deSalas:2017kay}, the authors did succeed in including data from DeepCore and ANTARES in their global fit, but similar problems with Super-K were reported.
Other groups have reported general difficulties, for example, in \cite{Capozzi:2025wyn} "For atmospheric neutrinos, currently involving hundreds of bins, dozens of systematic uncertainties, and refined statistical separation of event classes by flavor proxies, the construction of $\chi^2$ maps based on public information has become eventually unfeasible outside the experimental collaborations".
We understood this statement as a challenge rather than a warning.

\begin{figure}
    \centering
    \includegraphics[width=\linewidth]{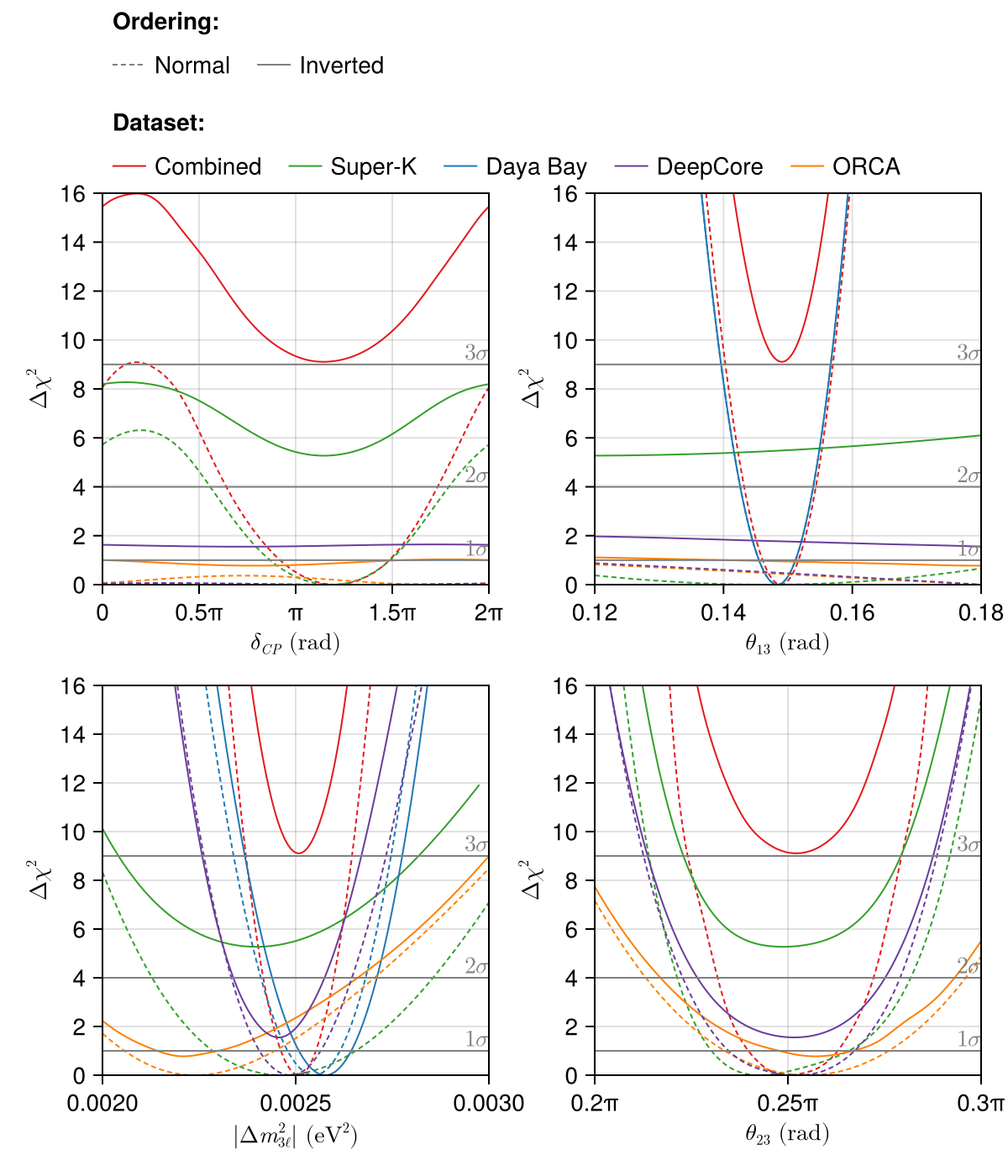}
    \caption{$\chi^2$ profiles for the atmospheric mixing parameters $|\Delta m^2_{3\ell}|$ and $\theta_{23}$, the leptonic CP violating phase $\delta_{CP}$, and the reactor angle $\theta_{13}$. Shown are separately the NO and IO profiles for the five analyses. Note that $\Delta m^2_{3\ell}$ corresponds to $\Delta m^2_{31}$ under NO and $\Delta m^2_{32}$ under IO.}
    \label{fig:global_results}
\end{figure}

At the time of writing, we did not find any combined analysis featuring more than one recent atmospheric dataset, and hence we present here the first of its kind, combining the data of Super-K, IceCube and KM3NeT, together with the Daya Bay reactor data, in a fully joint fit at the data level, including a careful modeling of the underlying physics with 91 fit parameters, fitting 839\,048 events distributed across 1536 bins simultaneously.
We find that a combined analysis can simultaneously describe all data well, and no large parameter tensions were observed. We report our measured values and uncertainties on the mixing parameters $\delta_{CP} = 3.78^{+0.89}_{-0.884}$, $\theta_{13}=0.149 ^{+0.00281}_{-0.00274}$, $\theta_{23}=0.785 ^{+0.0318}_{-0.0407}$, and $\Delta m_{31}^2=2.51 ^{+0.0463} _{-0.0441} \cdot10^{-3}\,\mathrm{eV}^2$, and find a preference of NO vs. IO at $\Delta\chi^2_{IO-NO}=9.11$, corresponding to $3\sigma$. We also disfavor a $\delta_{CP}$ value of zero at $\Delta\chi^2=8.06$. \Cref{fig:global_results} and \cref{tab:osc} give an overview of these main results.

\section{Datasets}
\label{sec:data}

\begin{figure}
    \centering
    \includegraphics[width=\linewidth]{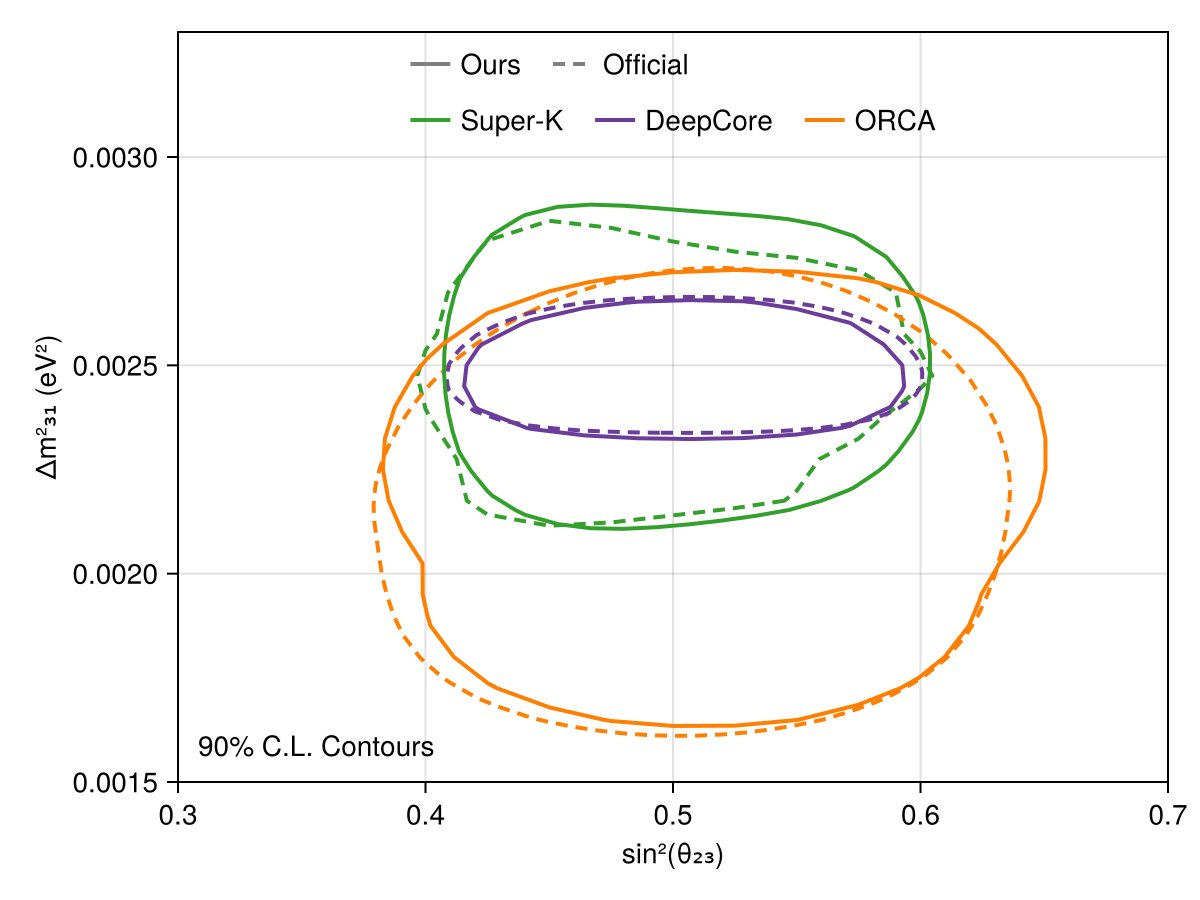}
    \caption{Validation plot for the three atmospheric datasets included. Shown are the 90\% C.L. contours for the atmospheric mixing parameters $\Delta m^2_{31}$ and $\sin^2(\theta_{23})$ for our implementation (solid lines) and the official results (dashed lines), for Normal Ordering.}
    \label{fig:2d}
\end{figure}

We introduce the publicly available datasets used for our analysis in the following, including a discussion of detector-specific uncertainties. The physics models (atmospheric neutrino flux, oscillation probabilities, and interaction cross sections) and analysis methods will be discussed later in Sec.~\ref{sec:analysis}.
In Fig.~\ref{fig:2d}, the standard 90\% C.L. 2d contours in the atmospheric mixing parameters plane $\Delta m^2_{31}$-$\sin^2(\theta_{23})$ can be found, comparing the values obtained from our implementations with the official reference curves. A similar validation plot for our Daya Bay implementation can be found in \cite{Kozynets:2024xgt}.

\subsection{Super-K}

We use Super-K phases I-V, which comprise 6511 days (484\,kton-years) of data. The official oscillation analysis of the sample was presented in 
\cite{Super-Kamiokande:2023ahc}, and more details can also be found in \cite{Wester:2023kac}. The analysis constrains the atmospheric mixing parameters $\Delta m^2_{31}$ and $\theta_{23}$, the leptonic CP violating phase $\delta_{CP}$, and the reactor angle $\theta_{13}$. The analysis reported a preference for the normal ordering at the $\approx 2\sigma$ level.

The Super-K collaboration provides a public dataset \cite{skdata} accompanying the publication, which we use here. The dataset provides expected and observed counts in all 930 analysis bins (expected is split up into $\nu_e$, $\bar{\nu}_e$, $\nu_\mu$, $\bar{\nu}_\mu$, $\nu_\tau + \bar{\nu}_\tau$ CC channels, and one flavor agnostic NC channel) for the best-fit NO and IO oscillation parameters, but without systematics applied.
The lack of systematic detector uncertainties is a limitation of the data release, and we approximate these by 38 separate fit parameters that handle bin migrations, selection efficiencies, normalizations, and energy scale uncertainties. The data release does not provide per-phase information, which means we cannot introduce per-phase uncertainties, as is done in the official analysis.
Another limitation of the dataset is that, for each analysis bin and channel, only summary statistics (mean, variance, and the [2.3, 5.9, 50, 84.1, 97.7]\,\% quantiles) for the bin's underlying true energy ($E_\nu$) and true zenith ($\theta$) distributions are provided. From this, we reconstruct a 2-dimensional probability distribution over the $(E_\nu, \cos{\theta})$ space by fitting a mixture of two log-normal distributions over the energy, and a von-Mises-Fisher distribution over $\cos{\theta}$. These distributions model the detector resolution, which effectively acts as a smearing matrix from true to reconstructed variables. To capture the energy-dependent correlation in the angular reconstruction resolution, we include an exponential energy dependence of the von Mises-Fisher $\kappa$ parameter, as motivated in \cite{Jesus-Valls:2025tfg}, as an additional per-bin fit parameter in the construction of the smearing matrices.

Since the data release provides separate expectations for NO and IO, this lets us test our reweighting procedure at nominal (i.e., without systematics applied) to check the smearing matrices. We get very good agreement for almost all bins (an average per-bin $\Delta\chi^2$ of 0.0012), except for some very low momentum bins: the FC SK I-III sub-GeV 1-ring $e$-like 0-decay electrons logP=[2.0, 2.6] (20 bins), the SK IV-V FC sub-GeV 1-ring $\bar{\nu}_e$-like 0-neutron logP=[2.0, 2.6] (20 bins), and the SK IV-V FC sub-GeV 1-ring $\bar{\nu}_e$-like 1-neutron logP=[2, 2.4] (10 bins). For these 50 bins, we get a per-bin $\Delta\chi^2$ of 0.017, which we consider too large. We suspect this is due to edge effects. These 50 offending bins do not contribute to the NMO signal (only 0.1\% of the total NMO difference is in these bins), and due to the observed mis-modeling, we exclude them from our analysis. This reduces the 930 bins of the original sample to 880 bins used here. If the actual resolution function were provided in a future data release, this situation could be avoided.

\begin{figure*}
    \centering
    \includegraphics[width=\linewidth]{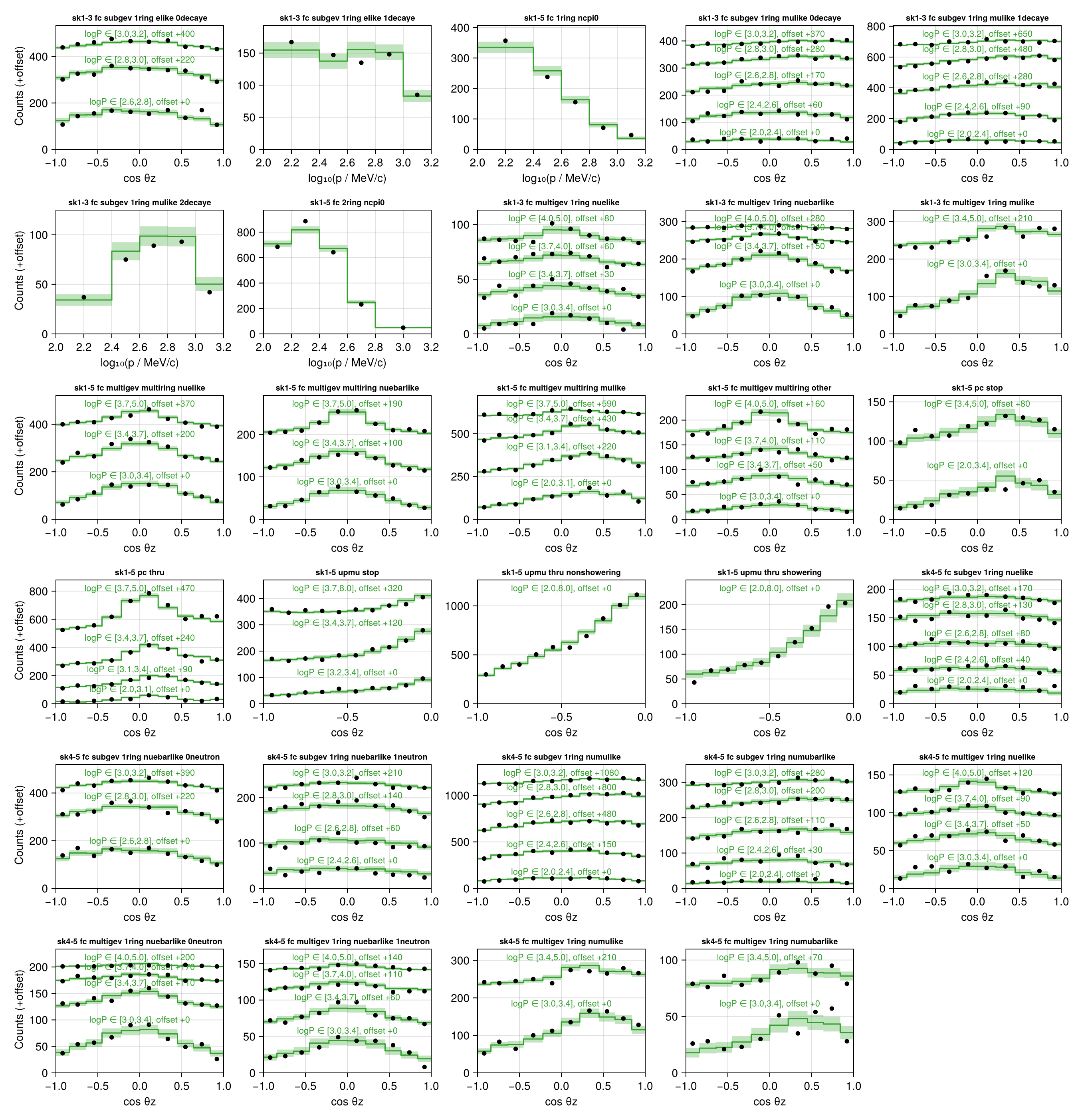}
    \caption{All Super-K data points (black) used in this analysis, overlayed with our model predictions (colorful bands) at the global best-fit point. The width of the bands represents the $1\sigma$ statistical uncertainty. The various reconstructed momentum (logP) bins are offset with constants provided in the plots to improve visibility.}
    \label{fig:sk_data}
\end{figure*}

\Cref{fig:sk_data} shows the Super-K data used together with the best-fitting model predictions (including systematic uncertainties). The overall goodness-of-fit when fitting Super-K alone is 15.9\%, and the preference for NO over IO is $\Delta\chi^2_{IO-NO}$ is 5.27 in excellent agreement with the official result of 5.23 (see \cref{tab:chi2}).

\subsection{DeepCore}

We use the IceCube-DeepCore analysis presented in \cite{IceCubeCollaboration:2023wtb}, which analyzes 7.5 years of effective lifetime of atmospheric neutrino data for the atmospheric mixing parameters $\Delta m^2_{31}$ and $\theta_{23}$.

The collaboration provides a public data release \cite{DVN/B4RITM_2025}, which we use here. The dataset provides Monte Carlo (MC) simulation and observed event counts in 200 analysis bins, split into 10 reconstructed energy bins spanning approximately 6.3 to 160\,GeV, 10 reconstructed zenith bins covering $\cos\theta \in [-1, 0.1]$, and 2 particle identification categories (Mixed and Tracks). The MC is provided separately for $\nu_e$ CC, $\nu_\mu$ CC, $\nu_\tau$ CC, and NC interactions for both neutrinos and antineutrinos, along with an atmospheric muon background contribution. Expected event counts are obtained by reweighting the MC to the oscillated atmospheric neutrino flux.

Detector systematic uncertainties are contained in the data release. These are modeled using precomputed linear response functions encoding the bin-by-bin sensitivities to bulk ice absorption, bulk ice scattering, overall DOM optical efficiency, and hole ice properties. These uncertainties are tabulated as a function of $|\Delta m^2_{31}|$ and interpolated linearly at each fit point to capture the mass-splitting dependence of the detection efficiency. Together with overall normalization (effective area scale) and atmospheric muon normalization, this results in seven detector-specific systematic parameters.

\begin{figure}
    \centering
    \includegraphics[width=\linewidth]{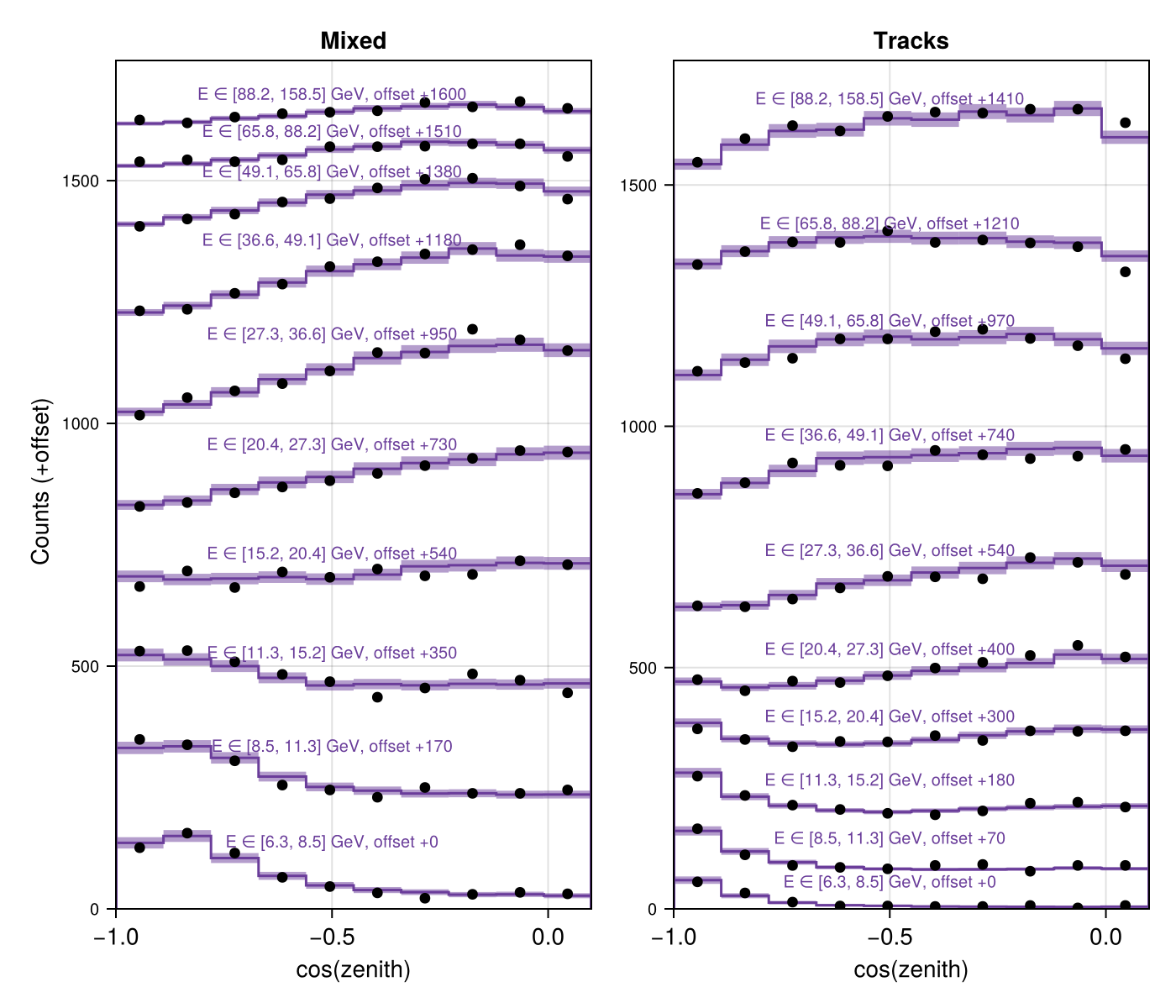}
    \caption{All DeepCore data points (black) used in this analysis, overlayed with our model predictions (colorful bands) at the global best-fit point. The width of the bands represents the $1\sigma$ statistical uncertainty. The various reconstructed energy bins are offset with constants provided in the plots to improve visibility.}
    \label{fig:dc_data}
\end{figure}

Figure \ref{fig:dc_data} shows the DeepCore data together with the best-fitting model predictions. The overall goodness-of-fit when fitting DeepCore alone is 28.4\% (see \cref{tab:chi2}), in good agreement with the 26.1\% reported in \cite{IceCubeCollaboration:2023wtb}. 

Our preference for NO over IO is $\Delta\chi^2_{IO-NO}=1.55$, while the collaboration does not provide this number. The PhD thesis \cite{maria_phd} reports a $\Delta\chi^2_{IO-NO}$ of 4.398 using a much larger DeepCore event sample with 9.3 years of data and 150\,257 events \cite{IceCubeCollaboration:2024ssx}, which is not available publicly.

\subsection{ORCA}

KM3NeT has published their first oscillation results using 510 days of data collected from a detector configuration with six lines (ORCA6) \cite{KM3NeT:2024ecf}, resulting in a 433 kton-years exposure. The data was used to measure both atmospheric mixing parameters and analyze differences between the two mass orderings. A dataset was made available with the publication \cite{FK2/Y0UXVW_2025}, which we use here.

The dataset provides binned response matrices to map from $(E_\nu, \cos{\theta_{true}})$ to $(E_{reco}, \cos{\theta_{reco}})$ per flavor, and separately for NC and CC interactions. Furthermore, the sample is divided into three event categories: "high-purity tracks, "low-purity tracks," and "showers". The dataset uses a very coarse grid of 10 logarithmically spaced points in true energy per decade. At these points, we reweight the spectra to account for changes in atmospheric flux, oscillations, and interaction cross-sections. To better capture the physics, we oversample $E_\nu$ to 50 points per decade, matching what we use for Super-K and DeepCore.
The data release does not include any parameterized systematic detector uncertainties, and the official analysis uses only an energy-scale uncertainty, which we also model. Otherwise, we implement the same five normalization uncertainties as in the official analysis, resulting in six detector-specific parameters.

\begin{figure}
    \centering
    \includegraphics[width=\linewidth]{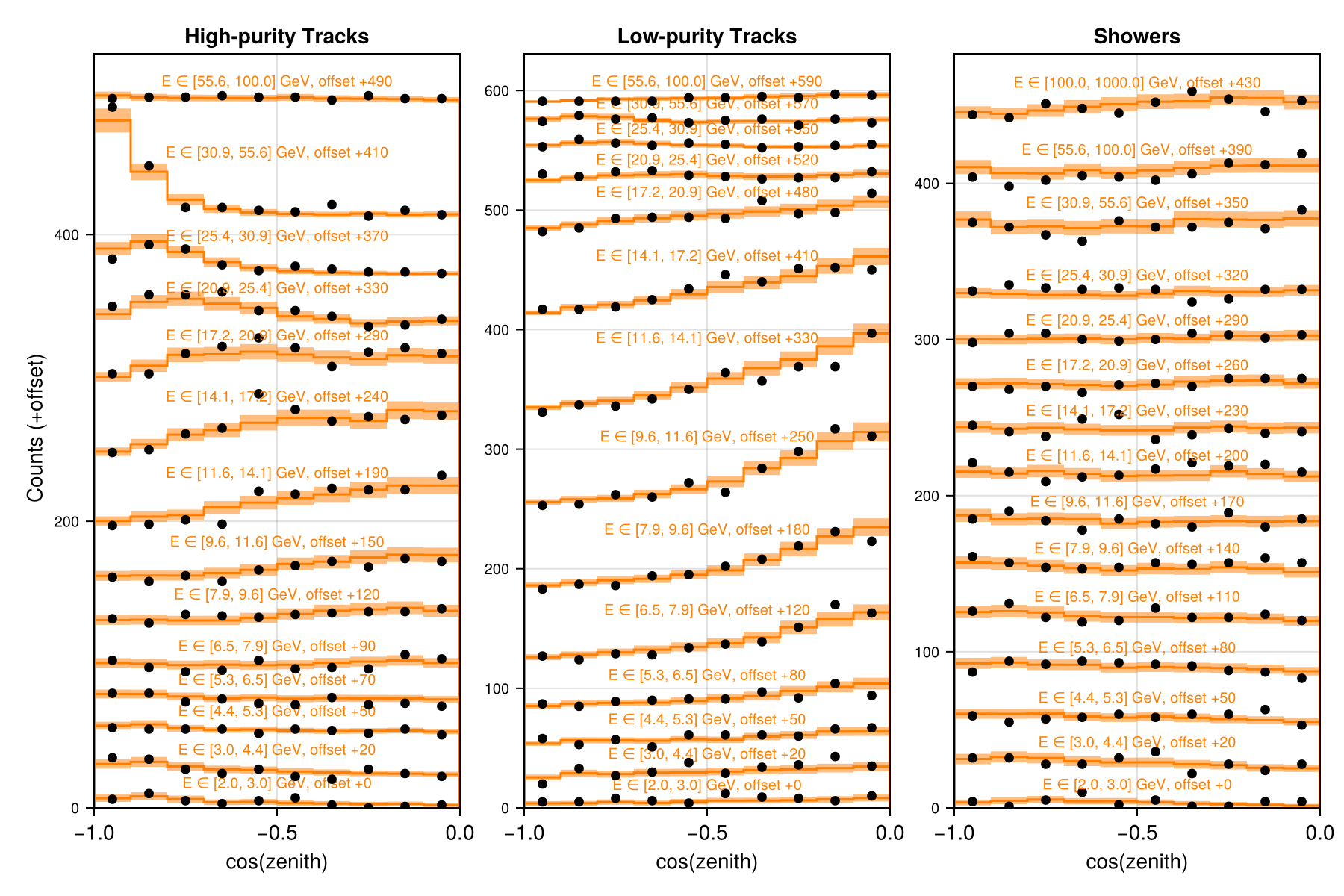}
    \caption{All ORCA data points (black) used in this analysis, overlayed with our model predictions (colorful bands) at the global best-fit point. The width of the bands represents the $1\sigma$ statistical uncertainty. The various reconstructed energy bins are offset with constants provided in the plots to improve visibility.}
    \label{fig:oc_data}
\end{figure}

The official g.o.f. p-value is reported to be 0.012 \cite{KM3NeT:2024ecf}, while we get 0.064 when fitting ORCA alone. This increased goodness-of-fit is largely due to the additional systematic uncertainties that we consider compared to the KM3NeT analysis  (see also \cref{tab:chi2}). When running with a reduced set of systematic uncertainties, similar to those used by the KM3NeT collaboration, our g.o.f. p-value reduces to 0.022.
The collaboration also reports a $\Delta\chi^2$ between NO and IO of 0.62, while we observe a compatible value of 0.78.
\Cref{fig:oc_data} shows the ORCA data with our best-fit predictions overlayed.

\subsection{Daya Bay}

Daya Bay was a reactor neutrino experiment that measured the reactor angle $\theta_{13}$ to world-leading precision and also constrained the atmospheric mass splitting \cite{DayaBay:2022orm}. The analysis is based on an event sample collected during 3158 days of operation, resulting in 752\,954 candidate neutrino events collected in the far detectors. 
The publication contains a data release as supplementary material, which we use here.
The data release provides a full covariance matrix to account for systematic uncertainties and their correlations between bins. This means we do not need any parameters other than the oscillation ones for Daya Bay in our fit. This treatment is justified, since the reactor neutrinos are fully complementary to the atmospheric neutrino data sets. The source of neutrinos is different, the neutrino energies are much lower (MeV), and the interactions happen via inverse beta decay. This means that there are no systematic uncertainties shared with the atmospheric experiments.

\begin{figure}
    \centering
    \includegraphics[width=\linewidth]{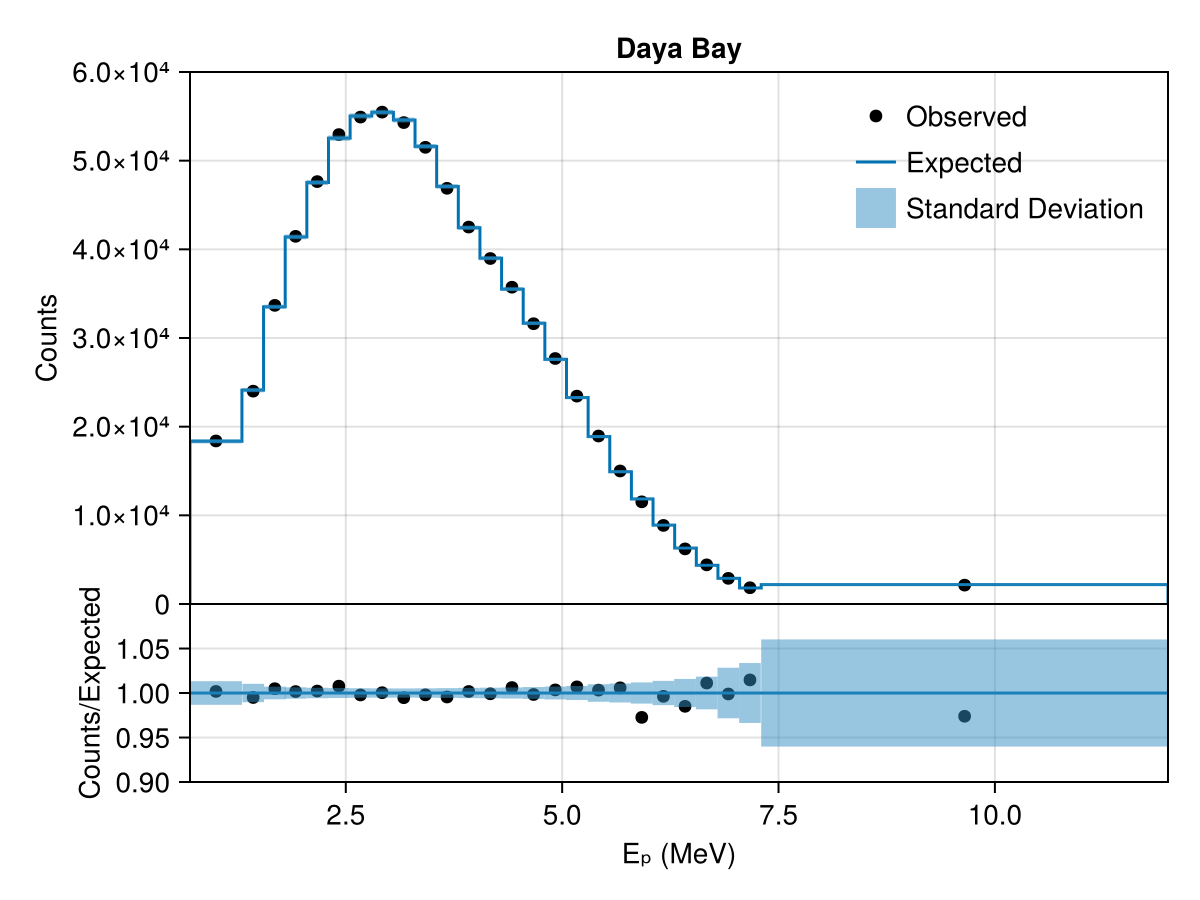}
    \caption{All Daya Bay data points (black) used in this analysis, overlayed with our model predictions (colorful bands) at the global best-fit point. The width of the bands represents the $1\sigma$ statistical and systematic uncertainty (from the covariance matrix).}
    \label{fig:db_data}
\end{figure}

\Cref{fig:db_data} shows the used data points in the 26 bins and our best-fitting expectation. We get an excellent $\chi^2$ goodness-of-fit of 58\%.
There is no preference ($\Delta\chi^2 < 10^{-4}$) for any mass ordering, as expected. See also \cref{tab:chi2}.

\section{Analysis}
\label{sec:analysis}

The analysis models the expected event rates from atmospheric neutrino fluxes, oscillation probabilities, and interaction cross sections. These effects are used to reweight the datasets provided by the collaborations. In addition, detector-specific uncertainties (as discussed in Sec.~\ref{sec:data}) are included, and a joint likelihood over the datasets is used, including a penalty term for priors on nuisance parameters, in a Frequentist analysis.

For the nominal predictions of the atmospheric neutrino fluxes, we rely on the calculations in \cite{Honda:2015fha}, which provide tabulated values for several locations \cite{HKKM}. We use the Kamioka site for Super-K, the South Pole site for DeepCore, and the Frejus site for ORCA.
The modeling of systematic uncertainties concerning flux ratios ($\nu_e/\bar{\nu}_e$, $\nu_\mu/\bar{\nu}_\mu$, and $\nu_e/\nu_\mu$) in three energy bands (sub-GeV, 1-10\,GeV, and above), and angular asymmetries (up-down and up-horizontal) follows \cite{Barr:2006it}, and an additional uncertainty $\Delta_\gamma$ modifying the spectral shape $E^{-(\gamma+\Delta_\gamma)}$. This totals to twelve parameters modifying the atmospheric flux included in our model.
The overall flux normalization, which is typically used by collaborations to account for other effects as well, is left unconstrained in the analysis. In ORCA and DeepCore, this is handled via a fit parameter for the overall normalization for each. In Super-K, it is handled via an overall norm, and two separate, energy-dependent parameters scaling the $<1\,$GeV and the $>1\,$GeV regions (see \cite{Wester:2023kac} for details).

We use the PREM model \cite{DZIEWONSKI1981297} to compute matter densities across 4 concentric Earth layers (inner core, outer core, mantle, crust). In addition, we add a 20\,km offset for the production height in the atmosphere. A fit parameter changes the electron density in the Earth around the nominal PREM values with a 6.8\% uncertainty.

To compute oscillation probabilities, we use the method presented in \cite{Maltoni:2023cpv}, which allows for correct averaging over energies and baselines in the presence of matter effects. We use this to average effects finer than our computational grid (50 points, log-spaced per decade of $E_\nu$), and to account for an uncertainty in the atmospheric production height of $\pm 10$\,km.

The neutrino interactions in the detectors were simulated with NEUT 5.4.0 for Super-K, while GENIE v2 (\textsc{G00\_00a}) \cite{Andreopoulos:2015wxa} was used in DeepCore and ORCA. In a combined analysis, however, we need to treat the datasets on a common ground; therefore, we reweight the DeepCore and ORCA spectra from GENIE to NEUT expectations. An equally valid option would be to keep ORCA and DeepCore at the GENIE reference values, and reweight Super-K instead. We tested that the choice of this reference does not significantly affect the results presented in this study (changes of $\lesssim 0.2\sigma$).

For cross-section uncertainties, we evaluated the per-process differences between the predictions of NEUT 5.4.0 and GENIE 3 with different tunes (\textsc{G18\_02a}, \textsc{G18\_10a}, and \textsc{G21\_11a}). This results in a normalization, a shape, and a $\nu/\bar{\nu}$ ratio uncertainty per interaction process. For charged-current (CC) interactions, we differentiate between quasi-elastic (1p1h), meson-exchange current (2p2h), resonant pion production (1pi), deep-inelastic scattering (DIS), and all other processes (other).  In addition, for quasi-elastic scattering, we add a parameter controlling the $\nu_e/\nu_\mu$ ratio, and introduce a separate normalization for sub-GeV events. Finally, we have an additional 25\% uncertainty on the $\nu_\tau$ CC cross section. For neutral current interactions, we only consider an inclusive reweighting. This results in 21 cross-section specific nuisance parameters included in our fit.

All data releases separate between flavours, CC and NC interactions, and $\nu$ vs. $\bar{\nu}$ neutrinos (only in Super-K, NC and $\nu_\tau$ CC do not provide $\nu$ vs. $\bar{\nu}$ separation), which we use to reweight the differential cross sections $d\sigma/dE$ separately. In the DeepCore data release, each MC event also contains the information of the simulated CC interaction process, which we use for per-process reweighting.

Data analysis is implemented in Julia, and the code is publicly available under the MIT open-source license \cite{Newtrinos,Eller:2026icp}. It interfaces with BAT.jl \cite{Schulz:2021BAT} for running statistical inference.
The code is fully forward- and reverse-mode auto-differentiable, which allows for the computation of partial derivatives used here for gradient descent optimization, the Fisher information, and Jacobians for change of variables.
The same framework has previously been used for the results shown in \cite{Ettengruber:2024fcq, Kozynets:2024xgt, Eller:2025lsh, Lonardi:2026exa}.

The analysis of Daya Bay data uses a $\chi^2$ test statistic including the provided covariance matrix to account for systematic uncertainties. All other datasets are fitted with a Poisson likelihood and added prior pull terms:
\begin{equation}
    \chi^2 = \sum_{i\in bins} -2log \left( \frac{e^{-E_i}E_i^{O_i}}{O_i!} \right) + \sum_{j \in priors} \frac{(\nu_j - \mu_j)^2}{\sigma_j^2}
\end{equation}
where $E_i$ and $O_i$ are the expected and observed number of events in each bin $i$, and the second sum over $j$ acts as a prior penalty for nuisance parameters $\nu_j$ that have a normal prior with mean $\mu_j$ and standard deviation $\sigma_j$.

We minimize the $\chi^2$ over all nuisance parameters $\nu_j$ to find maximum likelihood estimators (MLEs) $\hat{\nu}_j$ using gradient descent. To constrain the mixing parameters, we compute the profile likelihood ratio over fixed grids of parameter values while fitting all other parameters. Setting all parameters to their global MLEs for both orderings gives the difference between IO and NO, denoted $\Delta\chi2_{IO-NO}$ .

\section{Results}

We perform five separate analyses: fitting each experiment individually, and once in a combined fit. The combined fit needs to find parameter values that can accommodate all data at the same time. This means that one set of oscillation, flux, and cross-section parameters needs to be able to explain all 4 datasets, with a total of 839\,048 events divided into 1536 bins simultaneously. Because of the reduced degree of freedom in a combined fit, the goodness-of-fit per dataset will strictly degrade compared to that from the individual fits. The question is: will the combined fit still be able to explain the data, or will large tensions in shared parameters be observed? 

\Cref{tab:chi2} gives the breakdown of $\chi^2$ and goodness-of-fit p-values. The goodness-of-fit computed over individual datasets degrades from the individual to the combined fit only minimally, from 58\% to 55\% in the case of Daya Bay, as expected, since no nuisance parameters are shared. For Super-K, the number goes from 15.9\% to 14.2\%, and for DeepCore from 28.4\% to 21.1\%, which represent only modest decreases, and excellent goodness-of-fit was achieved in the combined case.
For ORCA, the goodness-of-fit was already not excellent in the individual fit (6.4\%), and in the combined case, it is at 1.1\%. This number is consistent with the official 1.2\% reported by the collaboration in \cite{KM3NeT:2024ecf}. The low ORCA p-value also causes the combined p-value across all datasets to be 2.3\%. This is a limitation of the ORCA dataset, and not our combined fit, as can be seen from removing the ORCA data points from the global g.o.f. computation, which increases the p-value to 14.1\%.

\begin{figure}[t]
    \centering
    \includegraphics[width=\linewidth]{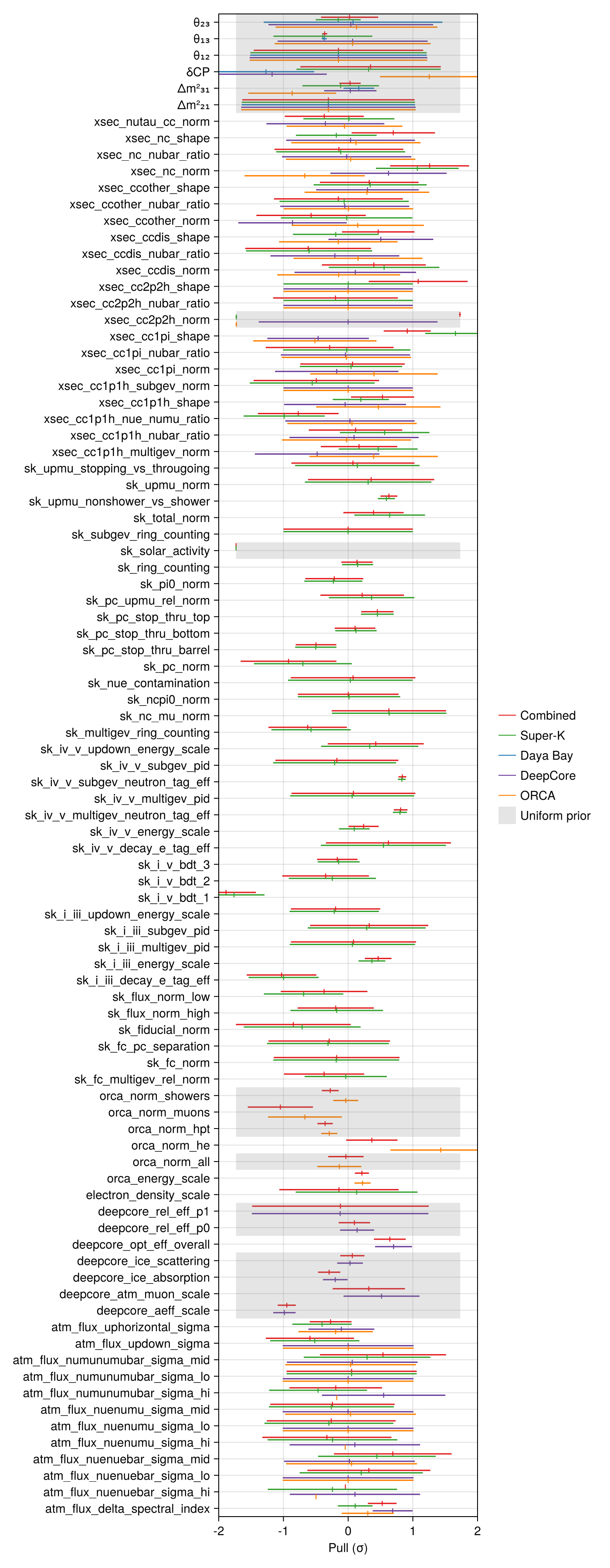}
    \caption{Best-fit parameter values and approximate uncertainty (see text) for the five different fits. The values are shown as pulls in the prior space. Uniform priors span $\pm \sqrt{12}/2$, indicated by gray boxes; otherwise, Normal priors are used.}
    \label{fig:pulls}
\end{figure}

\Cref{fig:pulls} provides further insight into the fit parameters under the five analyses. Shown are the best-fit values with uncertainties estimated via a Laplace approximation, where the posterior covariance is approximated by the inverse of the Fisher information matrix evaluated at the best-fit point. These uncertainties are approximate, assume Gaussian posteriors, and should not be directly compared with a proper profile-likelihood analysis. In particular, parameters with uniform priors whose best-fit values lie near the prior boundary are not well described by a Gaussian approximation and may have severely underestimated uncertainties (e.g. "xsec\_cc2p2h\_norm", "sk\_solar\_activity").
The plot demonstrates that there are no large differences in the fit parameters across the five analyses, indicating that the model can describe all datasets with the same underlying physics.

\begin{figure*}
    \centering
    \includegraphics[width=\linewidth]{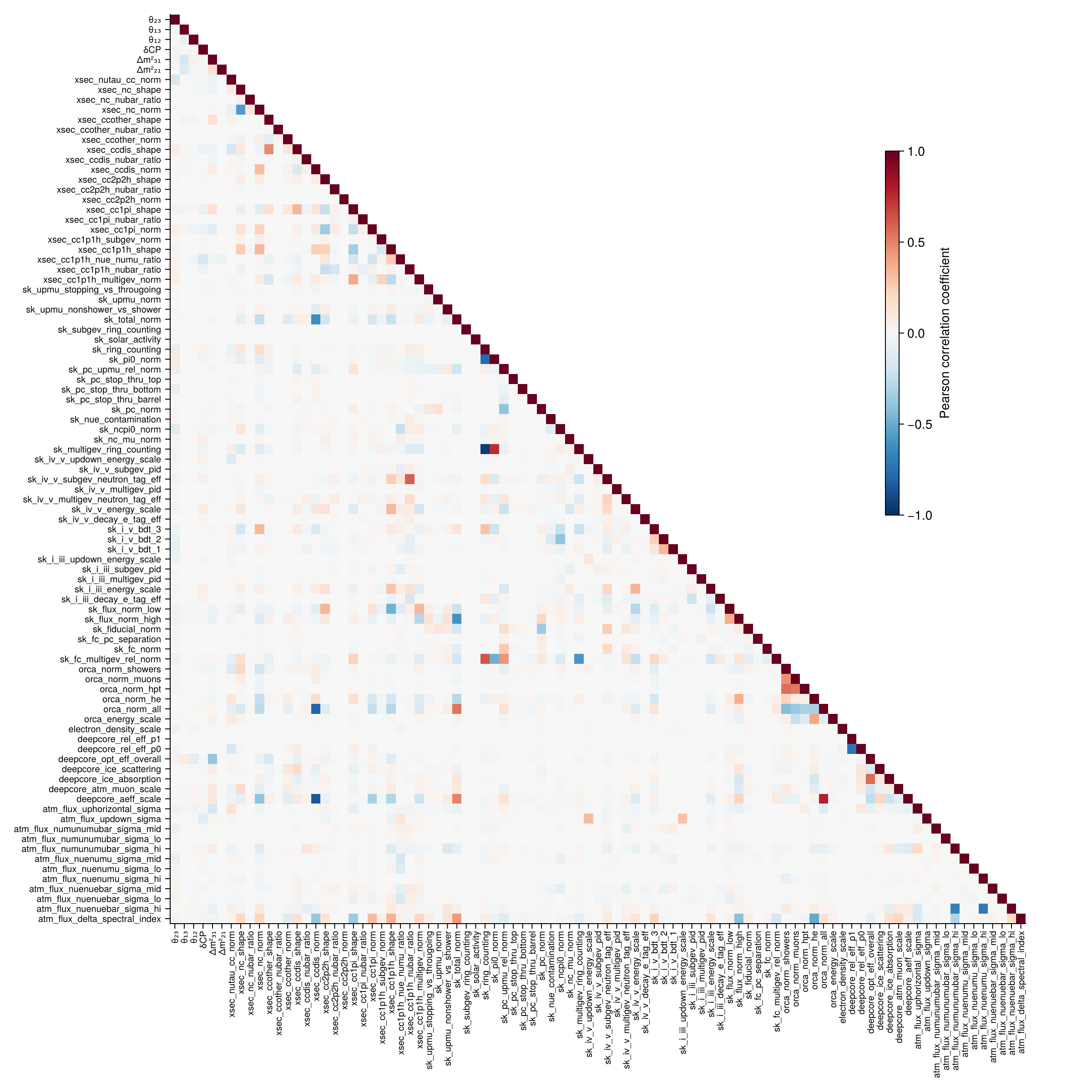}
    \caption{Pearson correlation coefficients of all fit parameters in the combined fit, obtained from inverting the Hessian (see text).}
    \label{fig:corr_mat}
\end{figure*}

\Cref{fig:corr_mat} provides further insight into the model parameters, showing the Pearson correlation coefficient (also obtained from the inverse Fisher information) for the combined fit. Some expected correlations are clearly visible, for example, the ORCA normalization parameters are strongly correlated with one another. So are the deepcore detector uncertainties. Furthermore, correlations can be seen, for example, between each experiment's overall normalization and cross-section normalizations.

\Cref{fig:global_results} shows the profile likelihood contours (converted to $\Delta\chi^2$) obtained from the five fits for the mixing parameters. As expected, Super-K is the only dataset with sensitivity to $\delta_{CP}$, and Daya Bay constrains $\theta_{13}$. All four experiments are sensitive to the atmospheric mass splitting. ORCA prefers a somewhat lower value, but has large uncertainties, the other experiments all fit around $2.5\cdot 10^{-3}$\,eV$^2$, and none of the fits are in tension. The atmospheric mixing angle is constrained by the three atmospheric datasets, all of which prefer maximal, or close to maximal, mixing. The best-fit numbers and uncertainties for the five analyses are reported in \cref{tab:osc}.

\begin{table}[tb]
\vspace{2em}
\begin{tabular}{l|lccc}
\textbf{Parameter}       & \textbf{Dataset}  & \textbf{Best fit} & \textbf{$+1 \sigma$} & \textbf{$-1 \sigma$} \\ \hline
\multirow{5}{*}{$\delta_{CP}$ (rad)} & Daya Bay           & -                & -                 & -                 \\
                         & Super-K           & 3.72             & 0.989             & 0.961             \\
                         & DeepCore          & -                & -                 & -                 \\
                         & ORCA              & -                & -                 & -                 \\
                         & \textbf{Combined} & \textbf{3.78}    & \textbf{0.89}     & \textbf{0.884}    \\ \hline
\multirow{5}{*}{$\theta_{13}$ (rad)} & Daya Bay           & 0.148            & 0.00286           & 0.00281           \\
                         & Super-K           & 0.147            & 0.046             & 0.0411            \\
                         & DeepCore          & -                & -                 & -                 \\
                         & ORCA              & -                & -                 & -                 \\
                         & \textbf{Combined} & \textbf{0.149}   & \textbf{0.00281}  & \textbf{0.00274}  \\ \hline
\multirow{5}{*}{$\Delta m_{31}^2$ ($10^{-3}$\,eV$^2$)}   & Daya Bay           & 2.55             & 0.0682            & 0.0672            \\
                         & Super-K           & 2.46             & 0.178             & 0.192             \\
                         & DeepCore          & 2.49             & 0.0803            & 0.0782            \\
                         & ORCA              & 2.24             & 0.182             & 0.207             \\
                         & \textbf{Combined} & \textbf{2.51}    & \textbf{0.0463}   & \textbf{0.0441}   \\ \hline
\multirow{5}{*}{$\theta_{23}$ (rad)} & Daya Bay           & -                & -                 & -                 \\
                         & Super-K           & 0.762            & 0.0388            & 0.0728            \\
                         & DeepCore          & 0.793            & 0.056             & 0.0463            \\
                         & ORCA              & 0.801            & 0.0672            & 0.0661            \\
                         & \textbf{Combined} & \textbf{0.785}   & \textbf{0.0318}   & \textbf{0.0407}   \\ \hline
\end{tabular}
\caption{Bestfit points and 68\% C.L. error-bars for the mixing parameters (NO), reported for each of the five analyses separately.}
\label{tab:osc}
\end{table}

A preference for normal over inverted ordering is present in all three atmospheric datasets. In the combined analysis including Daya Bay, this preference is at a $\Delta\chi^2_{IO-NO} = 9.11$. Assuming normality, this corresponds to approximately $3\sigma$ preference for the normal ordering \cite{Blennow:2013oma}.

\begin{table*}[]
\begin{tabular}{cccccccc}
\multirow{2}{*}{\textbf{Dataset}} & \multirow{2}{*}{\textbf{Bins}} & \multirow{2}{*}{\textbf{Events}} & \multicolumn{2}{c}{\textbf{Individual Fit}} & \multicolumn{2}{c}{\textbf{Combined Fit}} & \multirow{2}{*}{$\mathbf{\Delta\chi^2_{IO-NO}}$} \\
                                  &                                &                                  & $\chi^2$                 & p-value              & $\chi^2$                & p-value             &                        \\ \hline
Daya Bay                           & 26                             & 752954                           & 23.93                & 0.580                & 24.46               & 0.550               & 0.00                   \\
Super-K                           & 880                            & 58352                            & 921.86               & 0.159                & 925.06              & 0.142               & 5.27                   \\
DeepCore                          & 200                            & 21914                            & 210.95               & 0.284                & 215.80              & 0.211               & 1.55                   \\
ORCA                              & 430                            & 5828                           & 475.47               & 0.064                & 500.06              & 0.011               & 0.78                   \\
Combined                          & 1536                           & 839048                           & -                    & -                    & 1648.22             & 0.023               & 9.11                  \\ \hline
\end{tabular}
\caption{Summary of the datasets used in the five fits. Reported are also Pearson $\chi^2$, including prior penalty terms. The goodness-of-fit p-values are obtained from the $\chi^2$ distribution. The "Individual Fit" column refers to fitting one dataset at a time, while the "Combined Fit" uses the same best-fit parameter values from the combined fit for all datasets. The last column shows the $\chi^2$ difference between the best-fitting NO and IO expectations.}
\label{tab:chi2}
\end{table*}

\section{Conclusions}

We presented the first comprehensive combined oscillation analysis of multiple atmospheric neutrino datasets. Our combination includes data from Super-K, IceCube, and KM3NeT, paired with Daya Bay, and demonstrates the feasibility of such an analysis. Our study demonstrates that the data can be combined with a unified physics model, and no large parameter tensions were observed. 
Atmospheric data offer a powerful way to measure mixing parameters and to discern the neutrino mass orderings. We found consistent mixing parameter values when fitting all datasets individually and jointly. We disfavor the Inverted Neutrino Mass Ordering at $\Delta\chi^2=9.11$.

The three detectors used to collect the atmospheric samples had recently undergone extensive hardware upgrades: Super-K is taking data with Gadolinium loading \cite{Super-Kamiokande:2024kcb}, IceCube has successfully installed the IceCube-Upgrade \cite{IceCubeCollaborationP:2025rpl} featuring five new strings, and ORCA by now comprises 38 detector lines.
This means exciting new analyses and results on atmospheric neutrino oscillations are on the horizon. In a combined fit, either based on public data as presented here or in an inter-collaboration effort, a decisive answer to the open question of the neutrino mass ordering seems attainable in the near future \cite{Arguelles:2022hrt}.

When combining atmospherics with additional reactor data from JUNO, rather than only Daya Bay, another boost in NMO sensitivity is expected \cite{IceCube-Gen2:2019fet, KM3NeT:2021rkn}, further advocating for combined analyses that include atmospheric neutrinos.

A more immediate and practical improvement would also come from a more complete Super-K data 
release, including per-phase information, detector resolution functions, and a description of detector uncertainties. Similarly, updated and improved public data release by IceCube and KM3NeT would benefit combined analyses.

\section*{Acknowledgements}
We thank the Super-K, Daya Bay, IceCube, and KM3NeT collaborations for providing public data releases. P.E. is a member of the IceCube collaboration and an observer of the KM3NeT collaboration. We would like to thank O. Schulz, T. Wester, D. Schultheiß, T. Kozynets, and A. Zander for useful discussions.
This work has been supported by the Deutsche Forschungsgemeinschaft (DFG, German Research Foundation) under Germany's Excellence Strategy – EXC-2094/2 – 390783311, and the SFB 1258 – 283604770.

\bibliography{apssamp}

@article{Super-Kamiokande:2023ahc,
    author = "Wester, T. and others",
    collaboration = "Super-Kamiokande",
    title = "{Atmospheric neutrino oscillation analysis with neutron tagging and an expanded fiducial volume in Super-Kamiokande I{\textendash}V}",
    eprint = "2311.05105",
    archivePrefix = "arXiv",
    primaryClass = "hep-ex",
    doi = "10.1103/PhysRevD.109.072014",
    journal = "Phys. Rev. D",
    volume = "109",
    number = "7",
    pages = "072014",
    year = "2024"
}

@dataset{skdata,
    author       = "{Super-Kamiokande Collaboration}",
    title        = "{Data release: Atmospheric neutrino oscillation analysis with neutron tagging and an expanded fiducial volume in Super-Kamiokande I-V}",
    year         = 2023,
    publisher    = {Zenodo},
    version      = {v1},
    doi          = {10.5281/zenodo.8401262},
    url          = {https://doi.org/10.5281/zenodo.8401262}
}

@phdthesis{Wester:2023kac,
    author = "Wester, Thomas",
    title = "{Discerning the Neutrino Mass Ordering using Atmospheric Neutrinos in Super-Kamiokande I-V}",
    school = "Boston University, Boston U.",
    year = "2023"
}

@article{KM3NeT:2024ecf,
    author = "Aiello, S. and others",
    collaboration = "KM3NeT",
    title = "{Measurement of neutrino oscillation parameters with the first six detection units of KM3NeT/ORCA}",
    eprint = "2408.07015",
    archivePrefix = "arXiv",
    primaryClass = "hep-ex",
    doi = "10.1007/JHEP10(2024)206",
    journal = "JHEP",
    volume = "10",
    pages = "206",
    year = "2024",
    note = "[Addendum: JHEP 10, 041 (2025)]"
}

@dataset{FK2/Y0UXVW_2025,
author = "{KM3NeT Collaboration}",
publisher = {KM3NeT Open Data},
title = {{Oscillation Analysis ORCA433kt-y / ORCA 6}},
year = {2025},
version = {V2},
doi = {10.5072/FK2/Y0UXVW},
url = {https://doi.org/10.5072/FK2/Y0UXVW}
}

@article{DayaBay:2022orm,
    author = "An, F. P. and others",
    collaboration = "Daya Bay",
    title = "{Precision Measurement of Reactor Antineutrino Oscillation at Kilometer-Scale Baselines by Daya Bay}",
    eprint = "2211.14988",
    archivePrefix = "arXiv",
    primaryClass = "hep-ex",
    doi = "10.1103/PhysRevLett.130.161802",
    journal = "Phys. Rev. Lett.",
    volume = "130",
    number = "16",
    pages = "161802",
    year = "2023"
}

@article{IceCubeCollaboration:2023wtb,
    author = "Abbasi, R. and others",
    collaboration = "IceCube",
    title = "{Measurement of atmospheric neutrino mixing with improved IceCube DeepCore calibration and data processing}",
    eprint = "2304.12236",
    archivePrefix = "arXiv",
    primaryClass = "hep-ex",
    doi = "10.1103/PhysRevD.108.012014",
    journal = "Phys. Rev. D",
    volume = "108",
    number = "1",
    pages = "012014",
    year = "2023"
}

@dataset{DVN/B4RITM_2025,
author = "{IceCube Collaboration}",
publisher = {Harvard Dataverse},
title = {{Replication Data for: Measurement of atmospheric neutrino mixing with improved IceCube DeepCore calibration and data processing}},
UNF = {UNF:6:EqPPIAlmbhWU7MUgQgQVCw==},
year = {2025},
version = {V1},
doi = {10.7910/DVN/B4RITM},
url = {https://doi.org/10.7910/DVN/B4RITM}
}

@article{Honda:2015fha,
    author = "Honda, M. and Sajjad Athar, M. and Kajita, T. and Kasahara, K. and Midorikawa, S.",
    title = "{Atmospheric neutrino flux calculation using the NRLMSISE-00 atmospheric model}",
    eprint = "1502.03916",
    archivePrefix = "arXiv",
    primaryClass = "astro-ph.HE",
    doi = "10.1103/PhysRevD.92.023004",
    journal = "Phys. Rev. D",
    volume = "92",
    number = "2",
    pages = "023004",
    year = "2015"
}

@dataset{HKKM,
    author = "Honda, M. and Sajjad Athar, M. and Kajita, T. and Kasahara, K. and Midorikawa, S.",
    title = "{Atmospheric Neutrino Flux Tables for One-Year-Average (HAKKM, 2014).}",
    url = {http://www-rccn.icrr.u-tokyo.ac.jp/mhonda/public/nflx2014},
    year=2014
}

@article{Barr:2006it,
    author = "Barr, G. D. and Gaisser, T. K. and Robbins, S. and Stanev, Todor",
    title = "{Uncertainties in Atmospheric Neutrino Fluxes}",
    eprint = "astro-ph/0611266",
    archivePrefix = "arXiv",
    doi = "10.1103/PhysRevD.74.094009",
    journal = "Phys. Rev. D",
    volume = "74",
    pages = "094009",
    year = "2006"
}

@article{Maltoni:2023cpv,
    author = "Maltoni, Michele",
    title = "{From ray to spray: augmenting amplitudes and taming fast oscillations in fully numerical neutrino codes}",
    eprint = "2308.00037",
    archivePrefix = "arXiv",
    primaryClass = "hep-ph",
    reportNumber = "IFT-UAM/CSIC-23-99",
    doi = "10.1007/JHEP11(2023)033",
    journal = "JHEP",
    volume = "11",
    pages = "033",
    year = "2023"
}

@software{Newtrinos,
author = "Eller, Philipp and others",
publisher = {GitHub},
title = "Newtrinos.jl",
year = 2026,
url = {https://github.com/philippeller/Newtrinos.jl}
}

@article{Eller:2026icp,
    author = "Eller, Philipp and Schulthei{\ss}, David",
    title = "{Newtrinos.jl: A Julia Package for Global Analysis of Neutrino Data}",
    eprint = "2608.12940",
    archivePrefix = "arXiv",
    primaryClass = "hep-ph",
    month = "8",
    year = "2026",
    journal=""
}

@article{Ettengruber:2024fcq,
    author = "Ettengruber, Manuel and Zander, Alan and Eller, Philipp",
    title = "{Testing the number of neutrino species with a global fit of neutrino data}",
    eprint = "2402.00490",
    archivePrefix = "arXiv",
    primaryClass = "hep-ph",
    doi = "10.1103/PhysRevD.109.095016",
    journal = "Phys. Rev. D",
    volume = "109",
    number = "9",
    pages = "095016",
    year = "2024"
}

@article{Eller:2025lsh,
    author = "Eller, Philipp and Ettengruber, Manuel and Zander, Alan",
    title = "{Neutrino data analysis of extra-dimensional theories with massive bulk fields}",
    eprint = "2508.04274",
    archivePrefix = "arXiv",
    primaryClass = "hep-ph",
    doi = "10.1103/1llm-96vy",
    journal = "Phys. Rev. D",
    volume = "112",
    number = "5",
    pages = "055009",
    year = "2025"
}

@article{Kozynets:2024xgt,
    author = "Kozynets, Tetiana and Eller, Philipp and Zander, Alan and Ettengruber, Manuel and Koskinen, D. Jason",
    title = "{Constraints on non-unitary neutrino mixing in light of atmospheric and reactor neutrino data}",
    eprint = "2407.20388",
    archivePrefix = "arXiv",
    primaryClass = "hep-ph",
    doi = "10.1007/JHEP05(2025)130",
    journal = "JHEP",
    volume = "05",
    pages = "130",
    year = "2025"
}

@article{Lonardi:2026exa,
    author = "Lonardi, Sofia and Ettengruber, Manuel and Eller, Philipp",
    title = "{Searching for the $N$-naturalness tower of neutrinos}",
    eprint = "2607.14243",
    archivePrefix = "arXiv",
    primaryClass = "hep-ph",
    month = "7",
    year = "2026",
    journal=""

}

@article{Jesus-Valls:2025tfg,
    author = "Jes{\'u}s-Valls, C.",
    title = "{Parametrizing the reconstruction performance of Super-Kamiokande in the Sub-GeV to TeV neutrino energy range}",
    eprint = "2505.20812",
    archivePrefix = "arXiv",
    primaryClass = "hep-ex",
    month = "5",
    year = "2025",
    journal=""
}

@article{DZIEWONSKI1981297,
title = {Preliminary reference Earth model},
journal = {Physics of the Earth and Planetary Interiors},
volume = {25},
number = {4},
pages = {297-356},
year = {1981},
issn = {0031-9201},
doi = {https://doi.org/10.1016/0031-9201(81)90046-7},
url = {https://www.sciencedirect.com/science/article/pii/0031920181900467},
author = {Adam M. Dziewonski and Don L. Anderson}
}

@phdthesis{maria_phd,
    author = "Maria Veronica Prado Rodriguez",
    title = "Measurement of the neutrino mass ordering with 9.28 years of icecube deepcore data",
    school = "University of Wisconsin--Madison",
    year = "2024",
url="https://search.library.wisc.edu/catalog/9914259626302121"
}

@article{IceCubeCollaboration:2024ssx,
    author = "Abbasi, R. and others",
    collaboration = "(IceCube Collaboration){\ensuremath{\parallel}}, IceCube",
    title = "{Measurement of Atmospheric Neutrino Oscillation Parameters Using Convolutional Neural Networks with 9.3 Years of Data in IceCube DeepCore}",
    eprint = "2405.02163",
    archivePrefix = "arXiv",
    primaryClass = "hep-ex",
    doi = "10.1103/PhysRevLett.134.091801",
    journal = "Phys. Rev. Lett.",
    volume = "134",
    number = "9",
    pages = "091801",
    year = "2025"
}

@article{Andreopoulos:2015wxa,
    author = "Andreopoulos, Costas and Barry, Christopher and Dytman, Steve and Gallagher, Hugh and Golan, Tomasz and Hatcher, Robert and Perdue, Gabriel and Yarba, Julia",
    title = "{The GENIE Neutrino Monte Carlo Generator: Physics and User Manual}",
    eprint = "1510.05494",
    archivePrefix = "arXiv",
    primaryClass = "hep-ph",
    reportNumber = "FERMILAB-FN-1004-CD",
    month = "10",
    year = "2015",
    journal=""
}

@article{Super-Kamiokande:1998kpq,
    author = "Fukuda, Y. and others",
    collaboration = "Super-Kamiokande",
    title = "{Evidence for oscillation of atmospheric neutrinos}",
    eprint = "hep-ex/9807003",
    archivePrefix = "arXiv",
    reportNumber = "BU-98-17, ICRR-REPORT-422-98-18, UCI-98-8, KEK-PREPRINT-98-95, LSU-HEPA-5-98, UMD-98-003, SBHEP-98-5, TKU-PAP-98-06, TIT-HPE-98-09",
    doi = "10.1103/PhysRevLett.81.1562",
    journal = "Phys. Rev. Lett.",
    volume = "81",
    pages = "1562--1567",
    year = "1998"
}

@article{SNO:2002tuh,
    author = "Ahmad, Q. R. and others",
    collaboration = "SNO",
    title = "{Direct evidence for neutrino flavor transformation from neutral current interactions in the Sudbury Neutrino Observatory}",
    eprint = "nucl-ex/0204008",
    archivePrefix = "arXiv",
    doi = "10.1103/PhysRevLett.89.011301",
    journal = "Phys. Rev. Lett.",
    volume = "89",
    pages = "011301",
    year = "2002"
}

@article{T2K:2025yoy,
    author = "Abe, K. and others",
    collaboration = "T2K",
    title = "{Results from the T2K Experiment on Neutrino Mixing Including a New Far Detector {\ensuremath{\mu}}-like Sample}",
    eprint = "2506.05889",
    archivePrefix = "arXiv",
    primaryClass = "hep-ex",
    doi = "10.1103/gh5j-5cwv",
    journal = "Phys. Rev. Lett.",
    volume = "135",
    number = "26",
    pages = "261801",
    year = "2025"
}

@article{NOvA:2025tmb,
    author = "Abubakar, S. and others",
    collaboration = "NOvA",
    title = "{Precision Measurement of Neutrino Oscillation Parameters with 10 Years of Data from the NOvA Experiment}",
    eprint = "2509.04361",
    archivePrefix = "arXiv",
    primaryClass = "hep-ex",
    reportNumber = "FERMILAB-PUB-25-0619-PPD",
    doi = "10.1103/x53y-2b86",
    journal = "Phys. Rev. Lett.",
    volume = "136",
    number = "1",
    pages = "011802",
    year = "2026"
}

@article{Esteban:2020cvm,
    author = "Esteban, Ivan and Gonzalez-Garcia, M. C. and Maltoni, Michele and Schwetz, Thomas and Zhou, Albert",
    title = "{The fate of hints: updated global analysis of three-flavor neutrino oscillations}",
    eprint = "2007.14792",
    archivePrefix = "arXiv",
    primaryClass = "hep-ph",
    reportNumber = "IFT-UAM/CSIC-112, YITP-SB-2020-21",
    doi = "10.1007/JHEP09(2020)178",
    journal = "JHEP",
    volume = "09",
    pages = "178",
    year = "2020"
}

@article{T2K:2025wet,
    author = "Abubakar, S. and others",
    collaboration = "T2K, NOvA",
    title = "{Joint neutrino oscillation analysis from the T2K and NOvA experiments}",
    eprint = "2510.19888",
    archivePrefix = "arXiv",
    primaryClass = "hep-ex",
    reportNumber = "FERMILAB-PUB-25-0132-PPD",
    doi = "10.1038/s41586-025-09599-3",
    journal = "Nature",
    volume = "646",
    number = "8086",
    pages = "818--824",
    year = "2025"
}

@article{JUNO:2025gmd,
    author = "Abusleme, Angel and others",
    collaboration = "JUNO",
    title = "{Measurement of reactor neutrino oscillation with the first JUNO data}",
    eprint = "2511.14593",
    archivePrefix = "arXiv",
    primaryClass = "hep-ex",
    doi = "10.1038/s41586-026-10538-z",
    journal = "Nature",
    volume = "654",
    number = "8118",
    pages = "343--348",
    year = "2026"
}

@article{IceCube:2019dyb,
    author = "Aartsen, M. G. and others",
    collaboration = "IceCube",
    title = "{Development of an analysis to probe the neutrino mass ordering with atmospheric neutrinos using three years of IceCube DeepCore data}",
    eprint = "1902.07771",
    archivePrefix = "arXiv",
    primaryClass = "hep-ex",
    doi = "10.1140/epjc/s10052-019-7555-0",
    journal = "Eur. Phys. J. C",
    volume = "80",
    number = "1",
    pages = "9",
    year = "2020"
}

@article{Esteban:2016qun,
    author = "Esteban, Ivan and Gonzalez-Garcia, M. C. and Maltoni, Michele and Martinez-Soler, Ivan and Schwetz, Thomas",
    title = "{Updated fit to three neutrino mixing: exploring the accelerator-reactor complementarity}",
    eprint = "1611.01514",
    archivePrefix = "arXiv",
    primaryClass = "hep-ph",
    reportNumber = "IFT-UAM-CSIC-16-114, YITP-SB-16-45",
    doi = "10.1007/JHEP01(2017)087",
    journal = "JHEP",
    volume = "01",
    pages = "087",
    year = "2017"
}

@article{Gonzalez-Garcia:2014bfa,
    author = "Gonzalez-Garcia, M. C. and Maltoni, Michele and Schwetz, Thomas",
    title = "{Updated fit to three neutrino mixing: status of leptonic CP violation}",
    eprint = "1409.5439",
    archivePrefix = "arXiv",
    primaryClass = "hep-ph",
    reportNumber = "IFT-UAM-CSIC-14-095, YITP-SB-14-31",
    doi = "10.1007/JHEP11(2014)052",
    journal = "JHEP",
    volume = "11",
    pages = "052",
    year = "2014"
}

@article{Capozzi:2025wyn,
    author = "Capozzi, Francesco and Giar{\`e}, William and Lisi, Eligio and Marrone, Antonio and Melchiorri, Alessandro and Palazzo, Antonio",
    title = "{Neutrino masses and mixing: Entering the era of subpercent precision}",
    eprint = "2503.07752",
    archivePrefix = "arXiv",
    primaryClass = "hep-ph",
    doi = "10.1103/PhysRevD.111.093006",
    journal = "Phys. Rev. D",
    volume = "111",
    number = "9",
    pages = "093006",
    year = "2025"
}

@article{Esteban:2018azc,
    author = "Esteban, Ivan and Gonzalez-Garcia, M. C. and Hernandez-Cabezudo, Alvaro and Maltoni, Michele and Schwetz, Thomas",
    title = "{Global analysis of three-flavour neutrino oscillations: synergies and tensions in the determination of $\theta_{23}$, $\delta_{CP}$, and the mass ordering}",
    eprint = "1811.05487",
    archivePrefix = "arXiv",
    primaryClass = "hep-ph",
    reportNumber = "IFT-UAM/CSIC-18-112, YITP-SB-18-34",
    doi = "10.1007/JHEP01(2019)106",
    journal = "JHEP",
    volume = "01",
    pages = "106",
    year = "2019"
}

@article{Super-Kamiokande:2024kcb,
    author = "Abe, K. and others",
    collaboration = "Super-Kamiokande",
    title = "{Second gadolinium loading to Super-Kamiokande}",
    eprint = "2403.07796",
    archivePrefix = "arXiv",
    primaryClass = "physics.ins-det",
    doi = "10.1016/j.nima.2024.169480",
    journal = "Nucl. Instrum. Meth. A",
    volume = "1065",
    pages = "169480",
    year = "2024"
}

@article{IceCubeCollaborationP:2025rpl,
    author = "Abbasi, R. and others",
    collaboration = "IceCube",
    title = "{Physics potential of the IceCube Upgrade for atmospheric neutrino oscillations}",
    eprint = "2509.13066",
    archivePrefix = "arXiv",
    primaryClass = "hep-ex",
    doi = "10.1103/nnjw-jp1n",
    journal = "Phys. Rev. D",
    volume = "113",
    number = "7",
    pages = "072009",
    year = "2026"
}

@article{IceCube-Gen2:2019fet,
    author = "Aartsen, M. G. and others",
    collaboration = "IceCube-Gen2",
    title = "{Combined sensitivity to the neutrino mass ordering with JUNO, the IceCube Upgrade, and PINGU}",
    eprint = "1911.06745",
    archivePrefix = "arXiv",
    primaryClass = "hep-ex",
    doi = "10.1103/PhysRevD.101.032006",
    journal = "Phys. Rev. D",
    volume = "101",
    number = "3",
    pages = "032006",
    year = "2020"
}

@article{KM3NeT:2021rkn,
    author = "Aiello, S. and others",
    collaboration = "KM3NeT, JUNO",
    title = "{Combined sensitivity of JUNO and KM3NeT/ORCA to the neutrino mass ordering}",
    eprint = "2108.06293",
    archivePrefix = "arXiv",
    primaryClass = "hep-ex",
    doi = "10.1007/JHEP03(2022)055",
    journal = "JHEP",
    volume = "03",
    pages = "055",
    year = "2022"
}

@article{Blennow:2013oma,
    author = "Blennow, Mattias and Coloma, Pilar and Huber, Patrick and Schwetz, Thomas",
    title = "{Quantifying the sensitivity of oscillation experiments to the neutrino mass ordering}",
    eprint = "1311.1822",
    archivePrefix = "arXiv",
    primaryClass = "hep-ph",
    doi = "10.1007/JHEP03(2014)028",
    journal = "JHEP",
    volume = "03",
    pages = "028",
    year = "2014"
}

@article{T2K:2024wfn,
    author = "Abe, K. and others",
    collaboration = "T2K, Super-Kamiokande",
    title = "{First Joint Oscillation Analysis of Super-Kamiokande Atmospheric and T2K Accelerator Neutrino Data}",
    eprint = "2405.12488",
    archivePrefix = "arXiv",
    primaryClass = "hep-ex",
    doi = "10.1103/PhysRevLett.134.011801",
    journal = "Phys. Rev. Lett.",
    volume = "134",
    number = "1",
    pages = "011801",
    year = "2025"
}

@article{Schulz:2021BAT,
  author  = {Schulz, Oliver and Beaujean, Frederik and Caldwell, Allen and Grunwald, Cornelius and Hafych, Vasyl and Kr{\"o}ninger, Kevin and Cagnina, Salvatore La and R{\"o}hrig, Lars and Shtembari, Lolian},
  journal = {SN Computer Science},
  title   = {BAT.jl: A Julia-Based Tool for Bayesian Inference},
  year    = {2021},
  issn    = {2661-8907},
  month   = {Apr},
  number  = {3},
  pages   = {210},
  volume  = {2},
  day     = {12},
  doi     = {10.1007/s42979-021-00626-4},
  url     = {https://doi.org/10.1007/s42979-021-00626-4},
}

@article{KamLAND:2002uet,
    author = "Eguchi, K. and others",
    collaboration = "KamLAND",
    title = "{First results from KamLAND: Evidence for reactor anti-neutrino disappearance}",
    eprint = "hep-ex/0212021",
    archivePrefix = "arXiv",
    doi = "10.1103/PhysRevLett.90.021802",
    journal = "Phys. Rev. Lett.",
    volume = "90",
    pages = "021802",
    year = "2003"
}

@article{Arguelles:2022hrt,
    author = {Arg{\"u}elles, C. A. and Fern{\'a}ndez, P. and Mart{\'\i}nez-Soler, I. and Jin, M.},
    title = "{Measuring Oscillations with a Million Atmospheric Neutrinos}",
    eprint = "2211.02666",
    archivePrefix = "arXiv",
    primaryClass = "hep-ph",
    doi = "10.1103/PhysRevX.13.041055",
    journal = "Phys. Rev. X",
    volume = "13",
    number = "4",
    pages = "041055",
    year = "2023"
}

@article{deSalas:2020pgw,
    author = "de Salas, P. F. and Forero, D. V. and Gariazzo, S. and Mart{\'\i}nez-Mirav{\'e}, P. and Mena, O. and Ternes, C. A. and T{\'o}rtola, M. and Valle, J. W. F.",
    title = "{2020 global reassessment of the neutrino oscillation picture}",
    eprint = "2006.11237",
    archivePrefix = "arXiv",
    primaryClass = "hep-ph",
    doi = "10.1007/JHEP02(2021)071",
    journal = "JHEP",
    volume = "02",
    pages = "071",
    year = "2021"
}

@article{deSalas:2017kay,
    author = "de Salas, P. F. and Forero, D. V. and Ternes, C. A. and Tortola, M. and Valle, J. W. F.",
    title = "{Status of neutrino oscillations 2018: 3$\sigma$ hint for normal mass ordering and improved CP sensitivity}",
    eprint = "1708.01186",
    archivePrefix = "arXiv",
    primaryClass = "hep-ph",
    doi = "10.1016/j.physletb.2018.06.019",
    journal = "Phys. Lett. B",
    volume = "782",
    pages = "633--640",
    year = "2018"
}

\end{document}